\documentclass[prb,aps,twocolumn,preprintnumbers,amsmath,amssymb]{revtex4-1}

\usepackage{graphicx}

\usepackage{dcolumn}
\usepackage{bm}

\begin{document}

\title{Magnetic fluctuations in single-layer FeSe}
\author{T.~Shishidou}
\email{shishido@uwm.edu}
\author{D.~F.~Agterberg}
\author{M.~Weinert}
\affiliation{%
Department of Physics, University of Wisconsin-Milwaukee, Milwaukee, WI 53201
}%
\date{\today}
\begin{abstract}
The paramagnetic nature of monolayer FeSe films is investigated via first-principles spin-spiral
calculations.  
Although the ($\pi$,$\pi$) collinear antiferromagnetic (CL-AFM) mode -- the prevailing spin fluctuation mode relevant to most iron-based superconductors -- is lowest in
energy, the spin-wave energy dispersion $E(\bm{q})$ is 
found to be extremely flat over a large region of the two-dimensional Brillouin zone centered
at the checkerboard antiferromagnetic (CB-AFM) $\bm{q}$=0 configuration, giving rise to a sharp
peak in the spin density of states.  Considering the paramagnetic state as an incoherent average over
spin-spiral states, we find that resulting electronic band states around the Fermi level closely
resemble the bands of the CB-AFM configuration -- not the CL-AFM one -- and thus providing a natural
explanation of the angle-resolved photoemission observations.  The presence of the SrTiO$_3$(001)
substrate, both with and without interfacial oxygen vacancies, is found to reduce the energy
difference between the CB-AFM and CL-AFM states and hence enhance the CB-AFM-like fluctuations.  
These results suggest that CB-AFM fluctuations play a more important role than previously thought.  
\end{abstract}
\maketitle

\begin{figure*}
\includegraphics[width=0.95\textwidth]{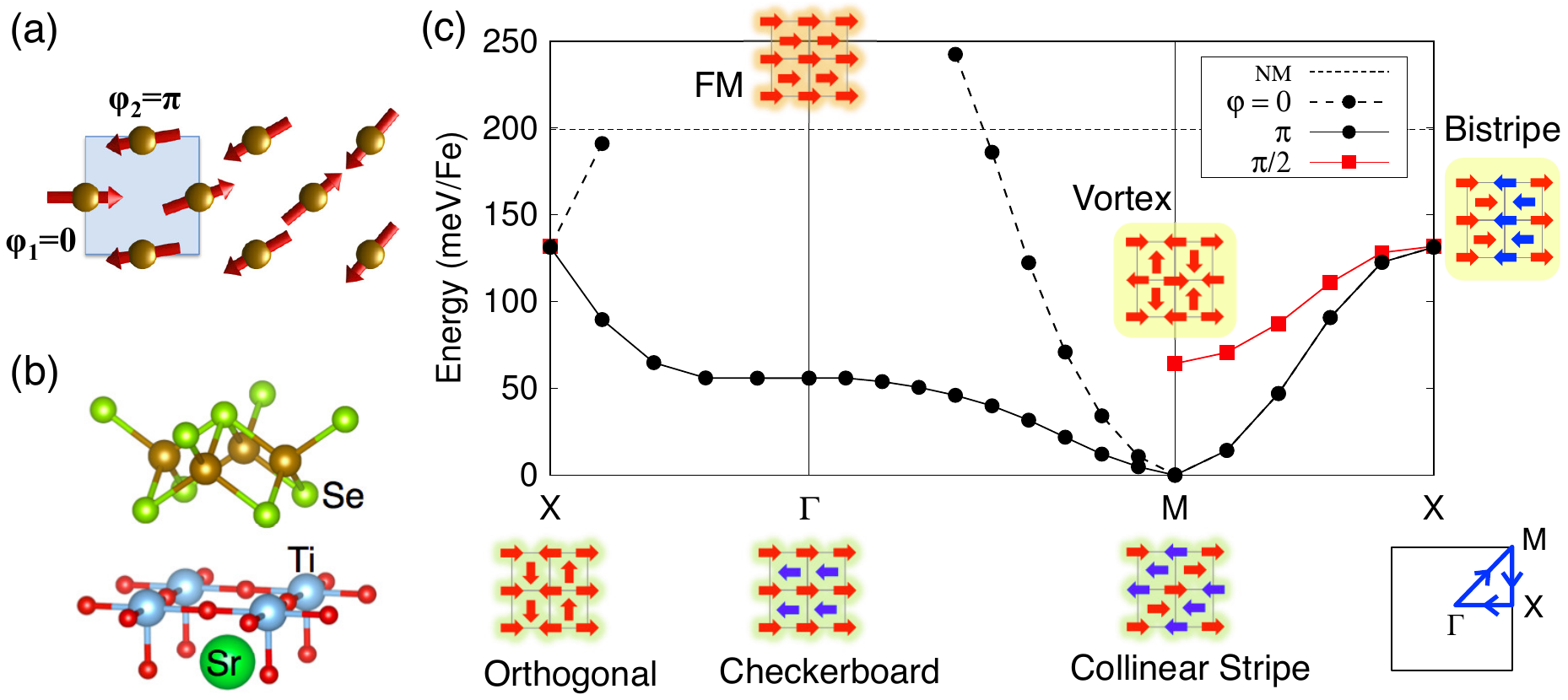}
\caption{\label{fig:Eq_free} 
\textbf{FeSe spin-spirals.}  
(a) Schematic of a planer spin-spiral with spin wave vector $\bm{q}=(1/36, 0)$ r.l.u.{} and relative atomic phase
$\varphi$=$\pi$.  Shaded square represents the crystallographic primitive cell.  
(b) Structural model of the FeSe monolayer on SrTiO$_3$(001) used in the present calculations; the bottom Se is
located above the surface Ti.   
(c) 
Spin-spiral total energies $E(\bm{q})$ per Fe of the freestanding monolayer FeSe film along high-symmetry
lines, relative to the collinear stripe energy. The horizontal dotted line is the energy of the non-magnetic
state.  Modes with relative atomic phases $\varphi$=0, $\pi$, $\pi/2$ are shown.   
Insets: Schematics of representative antiferromagnetic configurations at high-symmetry $\bm{q}$
points for $\varphi$=$\pi$ (green background) and $\pi/2$ (yellow background) modes.  
}
\end{figure*}

Single-layer FeSe films grown on SrTiO$_3$(001) (STO) have generated intense interest because of their reported
high superconducting critical temperature T$_\textrm{C}$$\sim$40--100
K.\cite{wang_interface,wen-hao,ge_superconductivity_2015, liu_electronic_2012, tan_interface-induced_2013,
he_phase_2013, liu_dichotomy_2014, lee_interfacial_2014, zhao_common_2016, wang_topological_2016} Angle-resolved
photoemission spectroscopy (ARPES) experiments\cite{liu_electronic_2012, tan_interface-induced_2013,
he_phase_2013, liu_dichotomy_2014, lee_interfacial_2014, zhao_common_2016, wang_topological_2016} reveal a number
of distinct features: As-prepared single-layer FeSe/STO films are not superconducting, but are
insulating or semiconducting.  After intensive vacuum annealing, however,  the monolayer films
are superconducting with a metallic Fermi surface.\cite{he_phase_2013,liu_dichotomy_2014}    
Contrary to most iron-based superconductors, this Fermi surface is characterized by electron pockets centered at
the Brillouin zone corner (M), with the zone center ($\Gamma$) states pushed below the Fermi level, posing a
challenge for pairing theories relying on Fermi surface nesting between the $\Gamma$ and M-centered
pockets\cite{Mazin_2008,Kuroki_2008} through ($\pi$,$\pi$) collinear stripe antiferromagnetic (CL-AFM) spin fluctuations
(c.f., Fig.~\ref{fig:Eq_free}(c) inset).   
Oxygen vacancies formed at the interface during the annealing process are
believed to play an important role in the transition to the metallic phase and high T$_\textrm{C}$ superconductivity.
Furthermore that bilayer FeSe films are not superconducting\cite{liu_dichotomy_2014} suggests that 
the FeSe-STO interface may play a key role in the superconducting mechanism.  

A number of first-principles density-functional theory (DFT) calculations have been reported for FeSe thin
films.\cite{wang_topological_2016,bang_2013,cao_2014,shanavas_doping_2015,xie_oxygen_2015,chen_effects_2016,
liu_2012_cal,bazhirov_2013_cal,zheng_antiferromagnetic_2013,cao_2015,liu_2015,zheng_potassium_2016}
Detailed comparisons of 
the experimental ARPES data\cite{wang_topological_2016}
for single-layer FeSe/STO 
to the DFT calculations corresponding to 
non-magnetic and various ordered antiferromagnetic configurations find that the calculated bands of only the
checkerboard antiferromagnetic (CB-AFM) configuration (Fig.~\ref{fig:Eq_free}(c) inset) are
consistent with the experimental data:
Both the Fermi surface and the band structure throughout the whole Brillouin zone (BZ) are reasonably
reproduced, particularly when a small Hubbard $U$ correction is included that pushes the
$\Gamma$-centered hole-like band completely below the Fermi level.  Other DFT studies have examined the effect of
interfacial oxygen vacancies
(O-vac):\cite{bang_2013,cao_2014,shanavas_doping_2015,xie_oxygen_2015,chen_effects_2016} 
oxygen vacancies are found to electron-dope the FeSe layer, and to modify the band structure of the
CB-AFM state, including the hole-like band around $\Gamma$ so that satisfactory agreement
with the ARPES data is achieved without the addition of a phenomenological $U$.\cite{chen_effects_2016}  

This seemingly successful agreement between the ARPES data and the DFT calculated bands for the CB-AFM
configuration of monolayer films of FeSe/STO, however is apparently inconsistent with the fact that the
CB-AFM configuration is not the calculated ground state. Rather, DFT total energy calculations consistently
find that the CL-AFM state is lower in
energy than the CB-AFM state for both monolayer FeSe and FeSe/STO, but the calculated
CL-AFM bands do not resemble the ARPES ones.  Although the energy difference between the CB- and CL-AFM
states is found to be reduced by electron doping due to the substrate, the CL-AFM state remains more stable
at reasonable doping levels.\cite{cao_2014,liu_2015,chen_effects_2016} Moreover, there is no experimental
evidence for {\it long-range\/} magnetic order in either bulk or monolayer FeSe  --- recent inelastic
neutron scattering experiments\cite{wang_magnetic_2016} have found that CB-AFM  and CL-AFM correlations
coexist even at low excitation energies in the bulk --- suggesting that a quantum paramagnetic state with
strong fluctuations between the CB-AFM and CL-AFM states might also exist in the monolayer system.
Furthermore, it has been suggested that a nematic paramagnetic state resembling the observed bulk
FeSe nematic state is a result of a near degeneracy between the CB-AFM and CL-AFM states.\cite{WaKL15}
Even in the case of coexisting CB-AFM and CL-AFM correlations, the question of why ARPES
experiments measure the CB-AFM-like band structure remains.

\begin{figure*}
\includegraphics[width=0.9\textwidth]{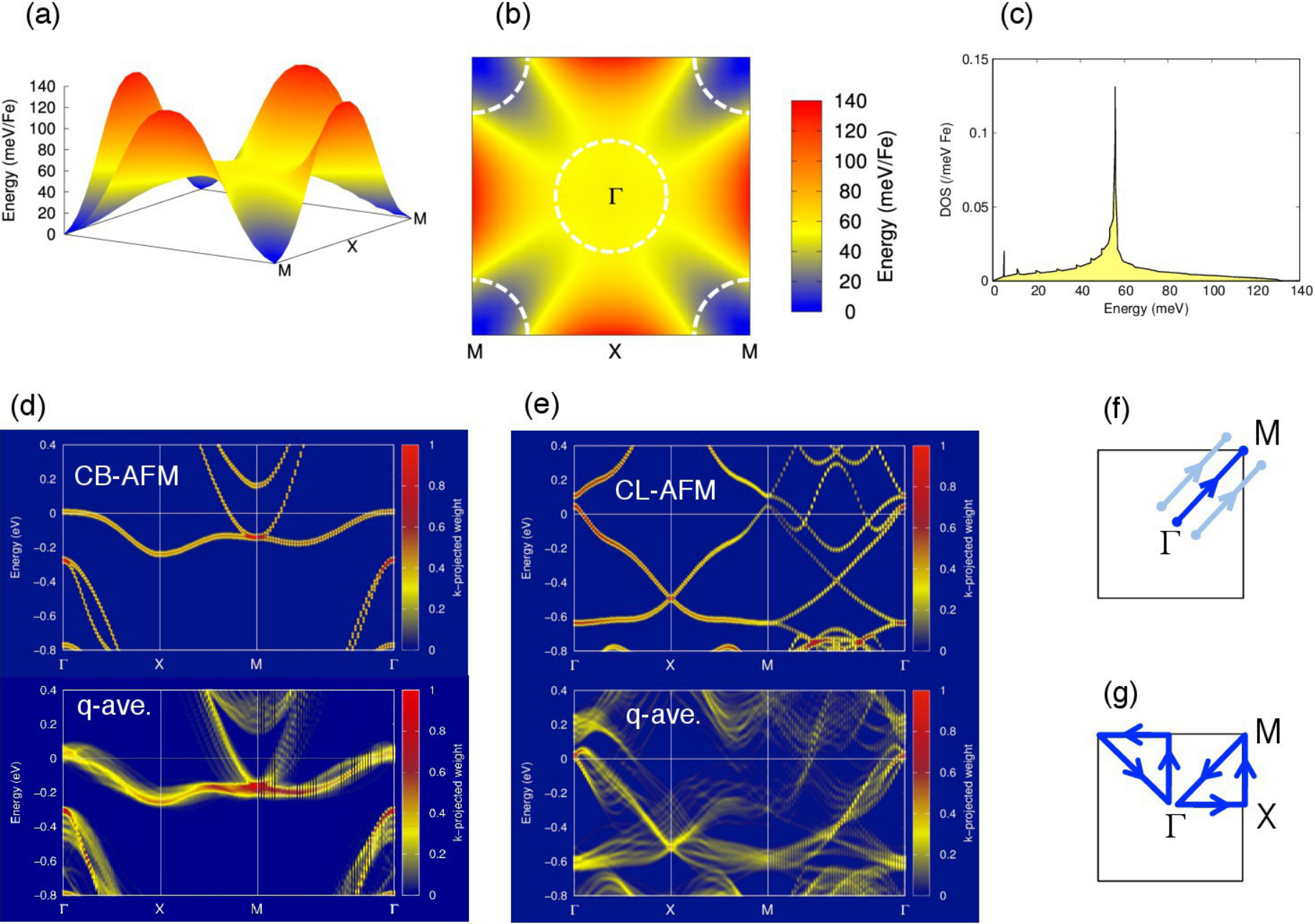}
\caption{\label{fig:EqBZ_free} 
\textbf{Spin-spiral energies and electronic spectral weight.}  
$E(\bm{q})$ $\varphi$=$\pi$ mode for freestanding FeSe over the first Brillouin zone (calculated on a
20$\times$20 mesh):  (a) Three-dimensional
landscape. (b) Two-dimensional map.  Dotted circles represent areas (radius 0.2 in units of $2\pi/a$) used in
calculating $q$-averaged spectral weight.  (c) Density of the $\pi$-mode spin-wave states.   
(d) Electronic energy band structure in the pure checkerboard antiferromagnetic configuration (top) and
$q$-averaged spectral weight (bottom) over the $\Gamma$-centered circle in (b). The Fermi level is set to zero.  
(e) Electronic bands in ideal collinear stripe configuration (top) and $q$-averaged spectral weight (bottom) over
the M-centered circle in (b).  
(f) Schematic showing the actual $k$-points at $\bm{k}^\pm = \bm{k}\mp \frac{\bm{q}}{2}$ (light blue) calculated
to get the dispersion along $\Gamma$--M (dark blue) for a spin-spiral of wave vector $\bm{q}$.    
(g) The two $\Gamma$--X--M--$\Gamma$ paths used to calculate the electronic spectral weights.
}
\end{figure*}

The apparent paramagnetic nature of FeSe implies (\textit{i}) there is no net magnetic moment or long-range
antiferromagnetic order, and (\textit{ii}) the translational (and space group) symmetry is consistent with the
crystallographic 4 atom unit cell. These conditions, however, do not imply or require that FeSe is non-magnetic.
Equating the paramagnetic state with the non-magnetic state leads to inconsistencies in the calculated properties
such as requiring a large band renormalization to match the ARPES data. Instead, the Hund's
rule tendency of Fe to form local moments and the existence of magnetic order in related Fe-based
superconducting materials, strongly indicate that inclusion of magnetic interactions and (local) moments are
essential for a proper description of the ground state properties. 

While ordered magnetic configurations can be calculated straightforwardly using standard DFT methods, treating the
paramagnetic state -- or disordered/random states in general --  is more difficult: Spin-dynamics DFT
calculations\cite{spin-dyn0,spin-dyn1} can, in principle, directly simulate the time evolution of the spins but
require both large supercells and long times, as well as being most applicable for configurations around an
ordered state. Coherent potential approximation-type approaches\cite{cpa,cpapm} effectively proceed by considering
averaged interactions.  Yet another approach is to model the paramagnetic state as an explicit configurational
average of a (large) number of different ``snap shots'' to mimic the various short-range interactions and
configurations that might occur.

We address the properties of magnetism in FeSe in the spirit of this last approach and show that spin fluctuations centered
around the CB-AFM configuration are unexpectedly important due to the large near degeneracy of spin wave states with
energies just above the CL-AFM state; 
our approach to the paramagnetic state provides the first natural resolution to the ARPES-DFT paradox.
The different magnetic
configurations are represented by spin-spiral states of spin-wave vector $\bm{q}$, each of which by
construction has no net magnetic moment but does have a local magnetic moment. The paramagnetic
state is then
represented as an incoherent sum of these states.  These individual planar spin-spiral states are calculated
by DFT non-collinear total energy calculations using the generalized Bloch theorem.\cite{Sandratskii_1991}
In Fig.~\ref{fig:Eq_free}(a), an example of a spin-spiral is shown for a spin configuration close to the
CB-AFM configuration, but with the Fe moments rotating slowly with spin wave vector $\bm{q}$:  An Fe at
$\bm{R}$ has its moment pointing in a direction given by the azimuth angle
$\phi$=$\bm{q}$$\cdot$$\bm{R}$+$\varphi_\alpha$, where $\varphi_\alpha$ is an atomic phase for the $\alpha$
atom in the (magnetic) unit cell.  Representative magnetic states at high symmetry points for two-Fe
(primitive chemical cell) spin-spiral configurations are shown in insets of Fig.~\ref{fig:Eq_free}(c).  While
standard DFT approaches using supercells can sample these ``discrete'' states and compare their energies,
spin-spiral calculations for varying $\bm{q}$ provide information on the connections of one representative
state to another in the context of the total energy $E(\bm{q})$ and the electronic band structure
$\epsilon_{n\bm{k}} (\bm{q})$.      

\noindent
\textsf{\bf Results}\\
The monolayer FeSe/STO system is modeled by a slab consisting of Se-Fe$_2$-Se/TiO$_2$/SrO as shown in
Fig.~\ref{fig:Eq_free}(b).  Although this is a minimal representation of the substrate STO, our results
reasonably reproduce previous calculations that use thicker STO substrates.\cite{cao_2014,chen_effects_2016,
liu_2015,liu_2012_cal}   The planer lattice constant $a$ is set to the STO parameter $\sim$3.9\AA. The
vertical heights of Fe and Se above the TiO$_2$ are relaxed in the CB-AFM state (as summarized in 
Table~I of the Supplementary Information),
and then held fixed during the spin-spiral calculations. To isolate the interfacial
effects, we consider a free-standing FeSe monolayer using the same structural parameters. In addition, we
also calculate spin-spirals for bulk FeSe at the experimental lattice
parameters\cite{glasbrenner_effect_2015} in order to compare to the inelastic neutron scattering
experiment\cite{wang_magnetic_2016} and to address the possible correspondence with magnetic fluctuations in
single-layer FeSe/STO.    

We first consider the freestanding FeSe monolayer film.  Figure \ref{fig:Eq_free}(c) shows the spin-spiral
total energy $E(\bm{q})$ with $\bm{q}$ along high-symmetry lines in the BZ\@.   Since there are two Fe atoms
per cell, several spin-spiral ``modes'' appear, characterized by the relative atomic phase
$\varphi$=$\varphi_2$$-$$\varphi_1$.  A mode with $\varphi$=$\pi$ (solid line connecting the filled circles) 
stably exists throughout the BZ and forms a ``band'' of lowest energy.   This band includes the CB-AFM state
at $\Gamma$, the CL-AFM state at M, and the non-collinear orthogonal state\cite{Pasrija_2013} at X\@.  Around the minimum at $\bm{q}$=M, $E(\bm{q})$ is mostly parabolic.
In contrast, $E(\bm{q})$ around $\Gamma$ (CB-AFM state) is markedly flat, i.e., 
there is almost no energy cost to slightly rotate the Fe moments from the CB-AFM configuration
and there are many spin-wave states with (almost) the same energy as that of the CB-AFM one.

The dashed line represents the $\varphi$=0 mode. This mode is degenerate with the $\pi$ mode at M on
the zone boundary, but splits away and corresponds to a ferromagnetic configuration at $\Gamma$.  This mode,
however, is highly unstable in the sense that the Fe moments collapse immediately as $\bm{q}$ goes away from
the zone boundary (c.f., Fig.~2 of Supplementary Information) 
and a self-consistent ferromagnetic configuration does not exist.  The
horizontal dotted line represents the energy of the non-magnetic (NM) state.  When $E(\bm{q})$ approaches or
exceeds this energy, the spiral calculation of the $\varphi$=0 mode either converges to the NM state or
never converges to a self-consistent solution.  We also find $\pi/2$ and $3\pi/2$ modes (represented by
filled squares connected by the red line) that exist only at the zone boundary and are degenerate in energy.    At M,
these modes give the non-collinear vortex state,\cite{ohalloran_stabilizing_2017} and at X the bicollinear stripe state.  

The energy splitting of the half-integer and integer $\pi$ modes is very large, amounting to $\sim$64 meV/Fe
at M, strongly indicating that non-Heisenberg interactions are important.\cite{glasbrenner_effect_2015}
In particular, in a simple Heisenberg model, the integer and half-integer $\pi$ modes would be degenerate at
M.  The inclusion of the previously considered fourth-order bi-quadratic term,\cite{glasbrenner_effect_2015} $(S_i\cdot S_j)^2$, however,
is not sufficient; from extensive fits of $E(\bm{q})$, more general 4-spin terms\cite{Takahashi,TS_spin}  
are essential to account for the DFT results. This non-Heisenberg behavior has implications for the
stability of possible paramagnetic states, including the nematic paramagnetic phase suggested for
bulk FeSe.\cite{WaKL15}

A description of the paramagnetic state in terms of the spin-spiral states requires knowledge of $E(\bm{q})$
not only along the high symmetry directions, but throughout the BZ\@.  The energy landscape of the $\pi$
mode for the free-standing FeSe monolayer is shown in Figs.~\ref{fig:EqBZ_free}(a)-(c).   The region around
$\Gamma$ is found to be exceedingly flat: for the $\Gamma$-centered circle of radius 0.2 ($2\pi$/$a$) shown
in Fig.~\ref{fig:EqBZ_free}(b), the energies are in a window of $-$3 to 1 meV relative to the CB-AFM energy
$E_\textrm{CB}$=56 meV\@, and the flat (yellow) region covers a substantially wide area of the 2D BZ,
resulting in a sharp peak in the density of states of the $\pi$ mode, Fig.~\ref{fig:EqBZ_free}(c).  At high
temperatures, entropy considerations suggest that many spin-wave states around $\Gamma$ will be
excited, and may be frozen in as the temperature is lowered.

The question that then arises is how the electronic structure varies for these $\bm{q}$ states near $\Gamma$
compared to the CB-AFM state.  To obtain the electronic energy bands in a spin-wave $\bm{q}$ state,
it is necessary
to carry out a $k$-projection procedure\cite{qi_2010,davenport_1988} (see ``Method'' for details),
leading to an approximate ``spectral weight'' $A_{\bm{q}}(\bm{k}, \varepsilon)$ rather than a sharp band
dispersion $\varepsilon_{n\bm{k}}$. The spectral weight of a paramagnetic state approximated by a
$q$-average of spin-spiral states is shown
in Fig.~\ref{fig:EqBZ_free}(d) together with the pure CB-AFM band structure; the $q$-resolved
evolution of the bands is given in the Supplementary Information.  The bands in the pure CB-AFM state (upper
panel) are consistent with previous calculations: there is an electron pocket at M and a hole-like band
around $\Gamma$, which touches the Fermi level. (This band around $\Gamma$ is pushed below the Fermi level,
in agreement with experiment, when oxygen vacancies in the STO substrate are
included.\cite{chen_effects_2016}) The $q$-averaged spectral weight, averaged over the $\Gamma$-centered
circle given in (b), shows features very similar to the pure CB-AFM bands.  Although the CB-AFM spin
configuration itself is easily deformed by spin wave formation, CB-AFM-like band features are very robust.
These results thus provide a natural explanation for the ARPES observation that the electronic bands look
like those of the CB-AFM configuration: entropy effects lead to a paramagnetic state built up from spin-wave states
centered around $\Gamma$ (extending over a large part of $q$-space), and these approximately degenerate
$\bm{q}$ states have similar electronic bands, such that the spectral weight looks similar to the CB-AFM.

The objection to this scenario is that the CL-AFM phase still is lower in energy, and so
the predicted electronic bands should also show CL-AFM features, in contrast to the ARPES data.   
The band structure of the pure CL-AFM state (the top panel of Fig.~\ref{fig:EqBZ_free}(e))
has hole pockets at M and $\Gamma$, in addition to two bands
crossing the Fermi level along $\Gamma$--M,
in good agreement with previous calculations.
Taking a $q$-average over a circle of radius 0.2 (2$\pi$/$a$) centered at the BZ corner, the spectral
weight is drastically changed from the pure CL-AFM bands as shown in the bottom panel of
Fig.~\ref{fig:EqBZ_free}(e).    At $\bm{k}$=M, the convex (hole-like) and concave (electron-like) bands repel
each other, with the result that states around M are pushed far away from the Fermi level.
Unlike the CB-AFM case, the electronic bands of spin-spiral states with $\bm{q}$ near M are very sensitive to
small changes in magnetic configuration; thus, the superposition of spin-wave states to form the paramagnetic
phase results in electronic bands that do not resemble the ideal CL-AFM ones.
Moreover, even if there are contributions from the $\varphi$=$\pi$ states throughout the BZ, the electronic
bands around M close to the Fermi level would be dominated by the CB-AFM-like bands since those are in the
correct energy range and have large spectral weight. (Around $\Gamma$ near the Fermi level, the bands are
somewhat similar. There are, however, more significant differences in the electronic bands that distinguish
the different magnetic configurations at energies less than about $-$0.4 eV.)

\begin{figure}
\includegraphics[width=\columnwidth]{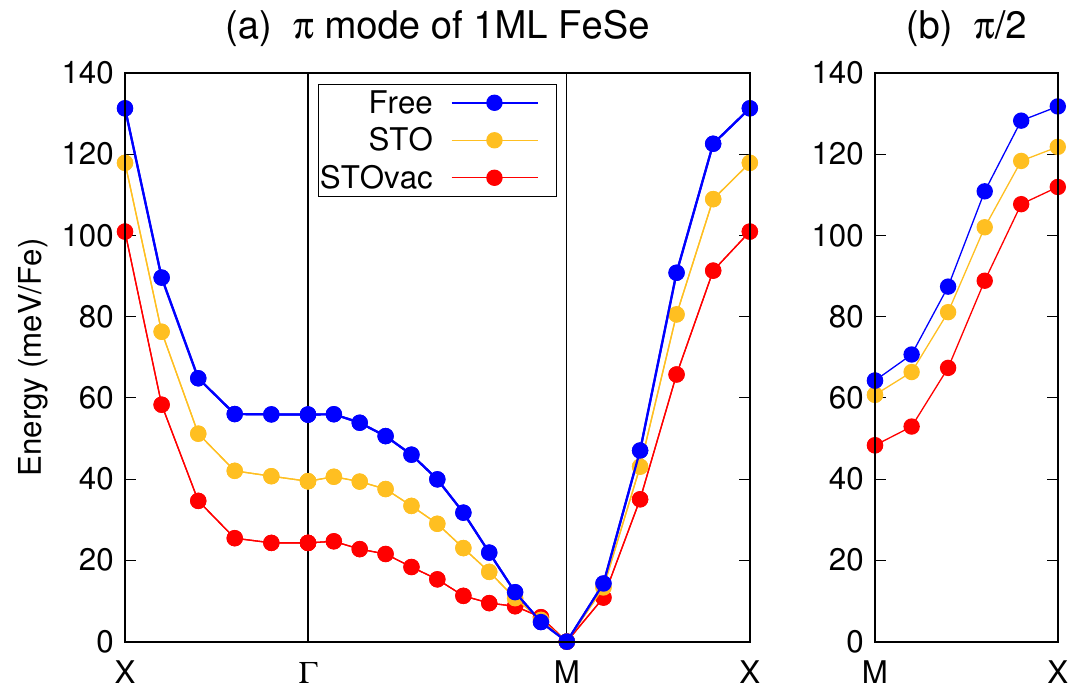}
\caption{\label{fig:Eq_STO_comp} 
\textbf{Comparison of spin-spiral energy dispersions.}  
$E(\bm{q})$ for monolayer freestanding FeSe (blue), FeSe/STO (yellow), FeSe/STO with interface oxygen vacancy (red)
modeled by setting $Z$=8.1.  (a) $\pi$ mode.   (b) $\pi/2$ mode.}
\end{figure}

Previous DFT calculations have shown that oxygen vacancies at the STO interface not only provide electron
doping to the FeSe layer, but also modify the electronic structure in the CB-AFM state significantly.
Self-developed electric fields cause spin splittings of the bands and open a gap for the $3d_{zx/zy}$
electron-like bands at M.\cite{zheng_antiferromagnetic_2013,chen_effects_2016} These effects have
been
attributed to hybridization with Se $4p_z$ state deformed by the electric
field.\cite{zheng_antiferromagnetic_2013} The $3d_{z^2}$ hole-like band around $\Gamma$, which touches the
Fermi level if a Coulomb correction $U$ is not added,\cite{wang_topological_2016} has its energy lowered
relative to the $3d_{zx/zy}$ band at M and its band width reduced, leading to a disappearance of the
$\Gamma$-centered hole pocket even without relying on $U$.\cite{chen_effects_2016}   
It is also reported that oxygen vacancies modify
the magnetic interaction and reduce the CB-AFM energy $E_\textrm{CB}$ relative to
CL-AFM\@.\cite{cao_2014,chen_effects_2016}  
Because of the possible importance of oxygen vacancies to the properties of FeSe/STO, we have also
considered their effect on $E(\bm{q})$.  
Here we use the conventional virtual crystal approximation without reducing the crystal symmetry.  
A single oxygen vacancy on the surface or at interface provides excess charge of nominally $\sim$$-2e$.   
A vacancy concentration $\alpha$ (0$\le\alpha\le$1) is simulated by replacing the atomic number of interface oxygen with $Z$=8+2$\alpha$.   
In Fig.~\ref{fig:Eq_STO_comp}, $E(\bm{q})$ for the $\pi$ modes of (\textit{i}) free-standing FeSe, and STO-supported
FeSe (\textit{ii}) with a perfect interface ($Z$=8) and (\textit{iii}) with oxygen vacancies ($Z$=8.1) are compared.
Although the flatness of $E(\bm{q})$ around $\Gamma$ remains in all cases, the CB-AFM energy $E_\textrm{CB}$
is significantly reduced by the presence of the STO substrate from 66 to 40 meV, and further reduced to 24
meV by the introduction of oxygen vacancies. However, the overall energy landscape still has the same shape as in
Fig.~\ref{fig:EqBZ_free}(a)-(c), and the variation of the electronic bands with $\bm{q}$ will be
similar. Thus, presence
of the STO substrate -- and oxygen vacancies -- will increase the propensity CB-AFM-like electronic
structure. 

For bulk FeSe we obtain a similar $E(\bm{q})$ landscape (Supplementary
Information): again the CL-AFM state is lower in energy than the CB-AFM state, and $E(\bm{q})$ is
flat around $\bm{q}$=$\Gamma$, and is consistent with
the experimentally observed coexistence of CB- and CL-AFM correlations in the bulk.\cite{wang_magnetic_2016} 
The similarity of $E(\bm{q})$ for bulk and films strongly support the idea that the same coexistence is
inherent  in FeSe/STO as well.  In addition, the increased flatness of bulk $E(\bm{q})$ around the
CL-AFM configuration ($\bm{q}$=M) implies that CL-AFM correlations are more important in the bulk
than the films while CB-AFM correlations are stronger in monolayer FeSe/STO\@.  These quantitative
differences in the bulk and film spin-spiral dispersions may be relevant to the difference in
superconducting T$_\textrm{C}$ and to the appearance of nematic order in the 3-D bulk.

\noindent
\textsf{\bf Conclusions}\\
The calculation of spin-spiral states provides a number of insights into the magnetic and electronic
properties of the monolayer FeSe, as well as providing an approach for consistently including (local)
magnetic effects in the treatment of the paramagnetic state.
Although the CB-AFM is not the calculated lowest energy ordered magnetic configuration, 
$E(\bm{q})$ is extremely flat around the $\Gamma$ point, resulting in a high density of states and entropy. These
almost degenerate spin-wave states have electronic bands similar to the pure CB-AFM state ($\bm{q}$=0). The
electronic bands for states around
low-energy CL-AFM configurations, on the other hand, are very sensitive to spin fluctuations, with the
electronic states near M are pushed away from the Fermi level. Thus for the paramagnetic phase described in terms
of spin-wave states, the resulting electronic structure around the Fermi level is expected to be dominated by
CB-AFM-like features, as observed in ARPES experiments. The STO substrate, including oxygen vacancies,
acts to reduce the energy of the CB-AFM ($\bm{q}$=0) vs.\ CL-AFM ($\bm{q}$=$(\pi,\pi)$), thus making both
CB-AFM-like magnetic correlations and electronic structure more favorable. Although we do not propose
any mechanism for the superconductivity, the magnetic fluctuations intrinsic to the paramagnetic state,
enhanced by proximity to the STO substrate, and the resulting effect on the description of the electronic
structure in the normal state are essential aspects that will need to be addressed in any comprehensive
theory of the superconductivity in the FeSe system; for example,
a recent effective $\bm{k}$$\cdot$$\bm{p}$-based theory
predicts\cite{Daniel} that the CB-AFM type fluctuations produces a fully-gapped nodeless $d$-wave
superconducting state on the M-centered electron pockets, naturally accounting for the gap anisotropy observed in single-layer FeSe films.\cite{PhysRevLett.117.117001}

\section*{Method}

All DFT calculations were carried out using the full-potential linearized augmented plane wave
method\cite{Mike_FLAPW_2009}  as implemented in the HiLAPW code, including noncollinear
magnetism\cite{PhysRevLett.111.176405} and generalized-Bloch-theorem spin-spiral magnetism.  Orientation of
the spin density is fully relaxed throughout the entire space.\cite{nakamura_2003}   The muffin-tin sphere
radius was set to 0.8\AA \ for oxygen and 1.1\AA \ for other atoms, and  the wave function and density and potential cutoffs were 16 and 200 Ry,
respectively.   The Perdew, Burke, and Ernzerhof form of the Generalized Gradient Approximation was used for
exchange correlation.  The Brillouin zone was sampled with 20$\times$20$\times$1 and 20$\times$20$\times$15
$k$-point meshes for the film and bulk calculations, respectively.  The density of two-dimensional spin-wave
states (Fig.~\ref{fig:EqBZ_free}(c)) was calculated by the triangle method.\cite{Lee_triangle_2002}  

In the spin-spiral calculations, the cell-periodic part $\tilde{\rho}$ of the spin off-diagonal density
matrix $\rho_{\uparrow\downarrow}$ for wave vector $\bm{q}$ has the form
$\tilde{\rho}=e^{i\bm{q}\cdot\bm{r}} \rho_{\uparrow\downarrow}$ in the interstitial region, and in the
sphere region $\tilde{\rho}=e^{i\bm{q}\cdot\bm{\tau}} \rho_{\uparrow\downarrow}$ with atomic position
$\bm{\tau}$.  

To obtain the band dispersion for a spin-wave state $\bm{q}$, a $k$-projection in the same spirit as the
unfolding technique for supercell calculations is needed.\cite{qi_2010,davenport_1988,chen_2014}     
The wave function of a band $\varepsilon_{n\bm{k}}(\bm{q})$ is a two component spinor, where each spin
component has a different Bloch phase,  
\begin{equation}
\psi_{n\bm{k}}^{\bm{q}} =  \begin{pmatrix}  
e^{i   (\bm{k}-\bm{q}/2)\cdot \bm{r} }  u_{n\bm{k}}^{\bm{q} \ \uparrow} \\
e^{i   (\bm{k}+\bm{q}/2)\cdot \bm{r} }  u_{n\bm{k}}^{\bm{q} \ \downarrow}  
\end{pmatrix} .
\end{equation}
Thus this band must be projected onto two different $k$ vectors,
\begin{equation}
\bm{k}^\pm = \bm{k} \mp \frac{\bm{q}}{2}  ,
\end{equation}
to obtain a band structure consistent with standard (supercell) calculations and with the momentum-resolved
measurement in ARPES experiments.  
The projection weights for $\bm{k}^\pm$ are $w^+=\langle u^\uparrow_{\bm{k}} | u^\uparrow_{\bm{k}} \rangle $ and $w^-=\langle u^\downarrow_{\bm{k}} | u^\downarrow_{\bm{k}} \rangle$.  
They can be found from the expectation value of $\sigma_z$ by 
\begin{equation}
w^\pm = \frac{1 \pm \langle \psi_{\bm{k}} | \sigma_z | \psi_{\bm{k}} \rangle }{2}
\end{equation}
since $\langle \psi_{\bm{k}} | \sigma_z | \psi_{\bm{k}} \rangle = w^+ - w^-$ and the orthonormality condition gives $w^+ + w^- = 1$.  
(The band index and $\bm{q}$ have been omitted for simplicity.)  The integrals implicit in the brackets
are performed over the primitive cell.  Figure~\ref{fig:EqBZ_free}(f) gives an example of how to
obtain the
dispersion along the $\Gamma$--M line: The band energies are calculated along
two $k$-rods shifted from the $\Gamma$--M segment by $\pm \bm{q}/2$, 
and then projected onto $\Gamma$--M.  To get the $q$-averaged spectral weights shown in
Fig.~\ref{fig:EqBZ_free}(d)(e), the two paths shown in (g) are averaged since they become inequivalent for arbitrary $\bm{q}$.

\section*{Acknowledgements}
This work was supported by the U.S. National Science Foundation,
Division of Materials Research, DMREF-1335215.

\section*{Author contributions}
TS and MW conceived the project and wrote the manuscript, TS carried out
the calculations, and all authors contributed to the interpretation of
the results and commented on the manuscript.

\section*{Competing financial interests}
The authors declare no competing financial interests.


\end{document}